\documentstyle[12pt,epsf]{article}
\newlength{\extraspace}
\newlength{\extraspaces}
\catcode`\@=11
\def\numberbysection{\@addtoreset{equation}{section}
\def\theequation{\arabic{section}.\arabic{equation}}}
\newcommand{\be}{\begin{equation}
\addtolength{\abovedisplayskip}{\extraspaces}
\addtolength{\belowdisplayskip}{\extraspaces}
\addtolength{\abovedisplayshortskip}{\extraspace}
\addtolength{\belowdisplayshortskip}{\extraspace}}
\newcommand{\ee}{\end{equation}}
\newcommand{\ba}{\begin{eqnarray}
\addtolength{\abovedisplayskip}{\extraspaces}
\addtolength{\belowdisplayskip}{\extraspaces}
\addtolength{\abovedisplayshortskip}{\extraspace}
\addtolength{\belowdisplayshortskip}{\extraspace}}
\newcommand{\ea}{\end{eqnarray}}
\newcommand{\newsection}[1]{
\vspace{7mm}
\pagebreak[3]
\addtocounter{section}{1}
\setcounter{equation}{0}
\setcounter{subsection}{0}
\setcounter{footnote}{0}
\begin{center}
{\large {\bf \thesection. #1}}
\end{center}
\nopagebreak
\medskip
\nopagebreak
\hspace{3mm}}

\newcommand{\tr}{\, {\rm tr}}
\newcommand{\e}{\, {\rm e}}

\newfont{\bg}{cmr10 scaled\magstep4}
\newcommand{\bigzerol}{\smash{\hbox{\bg 0}}}
\newcommand{\bigzerou}{\smash{\lower1.7ex\hbox{\bg 0}}}
\newcommand{\bighoshi}{\smash{\lower1.7ex\hbox{\bg *}}}

\begin{document}
\addtolength{\baselineskip}{.7mm}
\thispagestyle{empty}
\begin{flushright}
TIT/HEP-271 \\
NUP-A-94-19 \\
{\tt hep-th/9412215} \\
Dec, 1994
\end{flushright}
\vspace{2mm}
\begin{center}
{\Large{\bf Random Walk Construction of Spinor Fields \\
on Three Dimensional Lattice}} \\[15mm]
{\sc Masako Asano}\footnote{
\tt e-mail: maa@th.phys.titech.ac.jp}
,
{\sc Chigak Itoi}\footnote{
\tt e-mail: itoi@phys.cst.nihon-u.ac.jp}
and
{\sc Shin-ichi Kojima}\footnote{
\tt e-mail: kotori@th.phys.titech.ac.jp}

{\it $^{\ast \ddagger}$ Department of Physics, \\[2mm]
Tokyo Institute of Technology, \\[2mm]
Oh-okayama, Meguro, Tokyo 152, Japan} \\[4mm]
{\it $^{\dagger} $ Department of Physics, \\[2mm]
College of Science and Technology, Nihon University, \\[2mm]
Kanda, Surugadai, Chiyoda, Tokyo 101, Japan} \\[15mm]
{\bf Abstract}\\[5mm]
{\parbox{13cm}{\hspace{5mm}
Euclidean invariant Klein-Gordon, Dirac and massive Chern-Simons
field theories are constructed in terms of
a random walk with a spin factor on a three dimensional lattice.
We exactly calculate the free energy and the correlation functions
which allow us to obtain the critical diffusion constant and
associated critical exponents. It is pointed out that
these critical exponents do not satisfy the hyper-scaling
relation but the scaling inequalities.
We take the continuum limit of this theory on the basis
of these analyses.
We check the universality of obtained results on
other lattice structure such as triclinic lattice
and body centered lattice.
}}
\end{center}
\vfill
\newpage
\setcounter{section}{0}
\setcounter{equation}{0}
\numberbysection
%
%
%
\newsection{Introduction}
The random walk method has been studied extensively as
a useful tool in statistical mechanics as well as
in constructive field theory.
Many critical phenomena have been revealed
in this language.
For instance, in the two-dimensional Ising model,
the free energy and the correlation functions
are evaluated by summing over
all diagrams in the high temperature expansion
in terms of random walks with a spin factor \cite{V}.
This study also shows that Majorana spinor field can
describe critical phenomena.
As for $\varphi^4$ theory, the studies of Brownian motion
on an arbitrary dimensional lattice enables us to
prove important correlation inequalities in terms of a random
walk representation \cite{FFS}.
The construction in three-dimensions and the proof of triviality
in higher than four dimensions can be done using this method.
Although there have been many rigorous studies on
constructing a spinor field by
random walk method on a two dimensional lattice
and also on Brownian motion
on an arbitrary dimensional lattice for scalar field,
a random walk construction of a spinor field on a higher dimensional
lattice has so far never been obtained. \\

The purpose of this paper is to construct a theory of free spinor fields
by random walks on a three dimensional lattice.
We employ a regularization such that all quantities
can be evaluated as finite ones at any stage
before taking the limit of lattice spacing to be zero.
So far there have been some studies of a continuum random walk method for
spinor fields \cite{P}-\cite{IIM}.
Consider the following two point function of spinor fields
given in terms of a path integral representation \cite{IIM}:\\
\begin{equation}
< {\bf x}_f| \otimes<{\bf e}_f|
\frac{1}{J^{-1} J_{\mu} \partial_{\mu} + m}
|{\bf e}_i>\otimes |{\bf x}_i> =
\int \frac{{\cal D}{\bf x}}{V_{Diff}}
\exp \left[ -m \int^{1}_{0} dt \sqrt{\dot{{\bf x}}^2} +
iJ \Phi [\frac{\dot{{\bf x}}}{\sqrt{\dot{{\bf x}}^2}}
] \right],
\label{PI}
\end{equation}
where $J_{\mu}$ is spin matrix with the magnitude $J$
and $\exp[i J \Phi ]$ is the spin factor.
$\Phi[{\bf e}]$ is the oriented area determined by path
${\bf e}(t)$ between ${\bf e}_i={\bf e}(0)$ and ${\bf e}_f={\bf e}(1)$
on a two dimensional unit sphere \cite{IIM}.
The boundary condition of the path integral (\ref{PI}) requires initial
and final data
$$
{\bf x}_i \ ,   \ \ \ {\bf e}_i
= \frac{\dot{{\bf x}}_i}{\sqrt{{\dot{\bf x}_i}^2}} \ \ \ ;
\ \ \ \ \ \ \ \ \ {\bf x}_f \ ,
\ \ \  {\bf e}_f
= \frac{\dot{{\bf x}}_f}{\sqrt{{\dot{\bf x}_f}^2}} \ .
$$
The state vectors $<{\bf e}_f|$ and $|{\bf e}_i>$ are SU(2)
coherent states for the spin matrix $J_\mu$  \cite{Ch}.
In this paper, we give a mathematically well-defined
representation of the correlation function (\ref{PI})
with a lattice regularization.

First we work on a three dimensional regular cubic lattice.
We define a quantum diffusion process in terms of an amplitude with a
spin factor whose discrete time evolution obeys a certain recursion
relation.
This random walk model is a simple generalization of the diffusion
process for the two dimensional Ising model to the three dimensional model.
The free energy and the two point function are calculated by solving
the recursion relation.
We show that there are only $J=0, 1/2 , 1, 3/2$ and 2 cases on the
regular cubic lattice.
The critical diffusion constant for each $J$ is found at the singular
point of the free energy and its associated critical exponents are
obtained.
This analysis shows how we should scale the variables to obtain the
two point correlation function of the continuum field theory in our
random walk model.

We obtain Euclidean invariant field theories with spin 0, 1/2 and 1
for $J=0, 1/2$ and 1, respectively.
However, theories with spins higher than $1$ are not obtained from
this construction.
This is an unexpected result which cannot be understood in a naive
continuum random walk method for spinor fields.

The classification of universality
in $J=0, \ 1/2, \ 1 $ models
is nontrivial.
The order of phase transition of the model with $J= 0$ and 1/2 is
shown to be second and third, respectively.
In the $J=0$ case, the model belongs to the same universality class
as Brownian motion whose typical path has Hausdorff dimension 2.
The Klein-Gordon field is constructed at this critical point.
In $J=1/2$ and 1 case, a typical path has Hausdorff dimension 1.
The free Dirac field is constructed at the critical point of $J=1/2$
model.
It is worth noting that there is no fermion doubler in this model.
In the $J=1$ case, the critical behavior looks strange due to
the existence of Euclidean non-invariant excitations with long range
correlations in addition to an invariant one.
They contribute to the most singular terms of the free energy, which
suggests that the order of phase transition is first.
Nevertheless, we can take a suitable continuum limit so that
they do not appear in the correlation function.
The massive Chern-Simons field theory is constructed at this critical
point of $J=1$.
These non-invariant massless excitations affect critical exponents
which do not satisfy the hyper-scaling equality $d \nu = 2 - \alpha$
but {\it satisfies} the inequality $d \nu \geq 2 - \alpha$.

In the $J=3/2$ case, a rotational invariant correlation function
is never obtained.
In the $J=2$ case, there are two critical points for real $\kappa$,
one of them gives an Euclidean invariant scalar field theory
but the other does not.

We also investigate the lattice dependence of our model.
The universality of our construction should be checked on other lattice
structures.
We examine it on a triclinic lattice with bonds having identical length
and on a body centered lattice.
The results show that the spin 0 and
the spin 1/2 field theories are also obtained on these lattices
while the spin 1 field is not obtained in general.
\\

This paper is organized as follows.
In section 2, we give the definition of the random walk model
on a three dimensional cubic lattice.
We study the critical behavior of this random walk model in section 3.
In section 4, the Klein-Gordon, Dirac and massive Chern-Simons
fields are constructed by taking a scaling limit.
Various exponents are shown for convenience.
In section 5, we attempt to check the lattice universality of our results
obtained in the previous sections, especially on triclinic and body
centered lattices.
Some remarks and comments are given in section 6.

%
%
%
\newsection{3D Random Walks with Spin Factor}
In this section, we consider a random walk model on a
three dimensional cubic lattice ${\bf Z}^3$,
which is a discrete version of (\ref{PI})
\be
\log Z(\kappa, J) \sim \sum_{\omega} \kappa^{|\omega|}
\e^{iJ S(\omega)} \frac{1}{|\omega|} ,
\label{partitionfunction}
\ee
where $\sum_{\omega}$ denotes the sum over all closed
non-backtracking walks on the cubic lattice, $\kappa$ is a
diffusion constant and $J$ is a half-integer valued parameter
which will be interpreted later as {\it spin} of the model.
The quantity $S(\omega)$ is the oriented area encircled by
unit tangent vectors of a walk $\omega$ on $S^2$.
The weight for a path in this model is not
positive-real-valued but {\it complex-valued}.
It causes cancellation between walks, and the mean free distance
of random walks becomes longer than that for Brownian motion.
Such smoothing affects the critical behavior of our model.

\begin{figure}
    \begin{center}
     \leavevmode
       \epsfysize = 4cm
\hspace{.5mm}
       \epsfbox{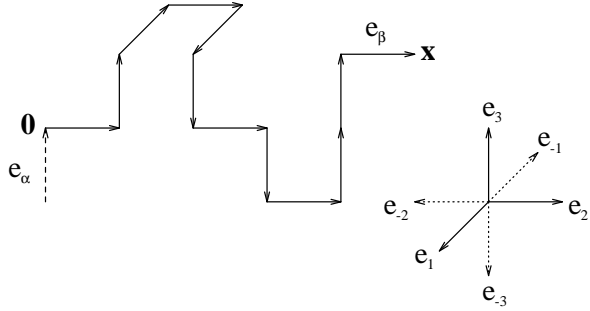}
    \end{center}
\caption{Figure of a walk start from the origin with the
         direction ${\bf e}_{\alpha}$ and come into a site ${\bf x}$
         from the direction ${\bf e}_{\beta}$ and the cooedinate
         system.}
\label{walk}
\end{figure}

Let us consider cubic lattice generated by three orthonormal vectors
${\bf e}_1 ,{\bf e}_2, {\bf e}_3$ which satisfy
${\bf e}_{\alpha}\cdot{\bf e}_{\beta} = \delta_{\alpha\beta} \;
(\alpha, \beta = \pm 1, \pm 2, \pm 3)$.
Here and after we use notation ${\bf e}_{-\alpha} = - {\bf e}_{\alpha}$.
This means the lattice unit is taken to be unity.
A site ${\bf x}$ is represented by $x^{\alpha} \in {\bf Z}$
as ${\bf x}= \sum^{3}_{\alpha=1} x^{\alpha}{\bf e}_{\alpha}$.
In order to sum up all possible walks, it is convenient to introduce
an amplitude $A^{\alpha\beta}_{l}({\bf x}| \kappa, J)$, which is
the sum of contributions from all walks which start from
the origin with the direction ${\bf e}_\alpha$ and
arrive at a site ${\bf x}$ from the direction ${\bf e}_\beta$
after $l$-steps Fig.\ref{walk}.

%
%

As we will see below, (\ref{partitionfunction})
can be represented on a lattice as a memory-1 walk.
The weight for a walk can be divided into two parts.
One is $\kappa^{|\omega|}/|\omega|$ which depends only on the
length of the walk $\omega$.
The other is the spin factor $\e^{iJS(\omega)}$
which gives non-zero phase when a walk curves.
To define such a phase on the lattice, we need successive
two steps of a discretized walk.
The amplitude obeys the following recursion relation
\be
  A^{\alpha\beta}_{l+1}({\bf x}| \kappa, J) = \kappa \sum_{\gamma \neq -\beta}
  \e^{iJS({\bf e}_3,{\bf e}_{\beta},{\bf e}_{\gamma})}
  A^{\alpha\gamma}_{l}({\bf x}-{\bf e}_{\beta}| \kappa, J) ,
  \label{recorsionrelation}
\ee
where
$-\alpha$ denote the direction of $-{\bf e}_\alpha$ and
$S({\bf e}_{\alpha},{\bf e}_{\beta},{\bf e}_{\gamma})$
is the area of a spherical triangle with vertices ${\bf e}_\alpha$,
${\bf e}_{\beta}$ and ${\bf e}_{\gamma}$ on $S^2$, basically.
The precise definition of the phase factor however,
will be given later.
The sum is taken over all nearest neighbour bonds except
the direction $-\beta$ in order to forbid backtracking processes.
The initial condition is given as
\be
A^{\alpha\beta}_{0}({\bf x}| \kappa, J) = \delta^{\alpha\beta}
       \delta_{x_1 0}\delta_{x_2 0}\delta_{x_3 0} .
\label{initialcondition}
\ee

Now we define the free energy density (free energy per unit volume)
in terms of the amplitude as
\be
f(\kappa, J) = -(-1)^{2J} \lim_{L \rightarrow \infty}
\sum_{l=1}^{L} \sum_{\alpha=\pm 1 }^{\pm 3} \frac{1}{l}
A^{\alpha\alpha}_{l}({\bf 0}| \kappa, J)  .
\label{freeenergy}
\ee
The factor $1/l$ in this definition removes $l$ degenerate
walks which represent the same configuration in the sum, as in
the case of two dimensional Ising model.
The factor $(-1)^{2J}$ attaches a minus-sign to every fermionic loop.
This is important to construct both bosonic
and fermionic theories in the same framework.

The free energy density eq.(\ref{freeenergy}) gives an explicit definition
of eq.(\ref{partitionfunction}).
The Fourier transformation of  the amplitude
$\tilde A^{\alpha\beta}_{l}({\bf p}| \kappa, J) = \sum_{{\bf x}}
A^{\alpha\beta}_{l}({\bf x}| \kappa, J) \e^{i{\bf p}\cdot{\bf x}}$
enables us to solve the recursion relation
\be
\tilde A^{\alpha\beta}_{l+1}({\bf p}| \kappa, J) =
\kappa \sum_{\gamma=\pm 1}^{\pm 3}
\tilde A^{\alpha\gamma}_{l}({\bf p}| \kappa, J)
Q^{\gamma\beta}({\bf p}, J) .
\label{fourierrecursion}
\ee
The matrix $Q^{\alpha\beta}({\bf p}, J)$ is a propagation matrix.
The elements of the matrix $Q^{\alpha\beta}({\bf p}, J)$ are given as follows:
\ba
Q({\bf p}, J) &=& Q({\bf 0}, J) \cdot P({\bf p})
\nonumber \\
\vspace{0.3mm}
\nonumber \\
Q({\bf 0}, J) &=& \left(
\matrix{
1 & e^{\frac{\pi}{2}iJ} & e^{2\pi iJ} & 0 & e^{-\frac{\pi}{2}iJ} & 1 \cr
e^{-\frac{\pi}{2}iJ} & 1 & e^{\pi iJ} & e^{\frac{\pi}{2} iJ} & 0 & 1 \cr
e^{-2\pi iJ} & e^{-\pi iJ} & 1 & 1 & e^{\pi iJ} & 0 \cr
0 & e^{-\frac{\pi}{2}iJ} & 1 & 1 & e^{\frac{\pi}{2}iJ} & 1 \cr
e^{\frac{\pi}{2}iJ} & 0 & e^{-\pi iJ} & e^{-\frac{\pi}{2}iJ} & 1 & 1 \cr
1 & 1 & 0 & 1 & 1 & 1 \cr}
\right) \label{PROQ}\\
P({\bf p}) &=& \left(
\matrix{
e^{i p_1} & & & & \bigzerou & \cr
 & e^{i p_2} & & & & \cr
 & & e^{i p_3} & & & \cr
 & & & e^{-i p_1} & & \cr
 & & & & e^{-i p_2} & \cr
 & \bigzerol & & & & e^{-i p_3} \cr
 }
\right) .
\nonumber
\ea
By (\ref{PROQ}), we give the definition of the spin factor.
Several comments must be noted.
The representation of the matrix $Q({\bf 0}, J)$ is not the unique
choice to obtain the same free energy density as (\ref{freeenergy}).
The zero components in the propagation matrix show the non-backtracking
condition.
The parameter $J$ {\it must} be chosen to be a half-integer
$$
J \in {\bf Z}/2
$$
in order that the total phase factor for any closed path produces
oriented area on $S^2$.
Note that the matrix $Q^{\alpha\beta}({\bf p}, J)$ is {\it hermitian}
at ${\bf p}={\bf 0}$.

%
\begin{figure}[tbhp]
    \begin{center}
     \leavevmode
       \epsfysize = 6cm
\hspace{.5mm}
       \epsfbox{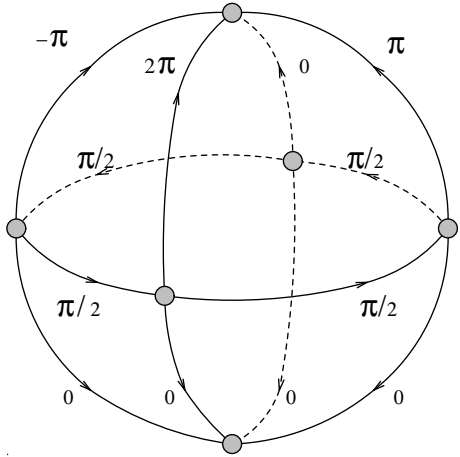}
    \end{center}
\caption{An assignment of phase-factor.}
\label{asanozu1}
\end{figure}

Using eq.(\ref{fourierrecursion}) and the condition (\ref{initialcondition})
the free energy density is reduced to the following form
\ba
f(\kappa, J) &=& -(-1)^{2J}\lim_{L\rightarrow\infty} \int^{\pi}_{-\pi}
\frac{d^3 p}{(2\pi)^3} \tr \sum^{L}_{l=1} \frac 1 l
[\kappa Q({\bf p}, J)]^l
\nonumber \\
&=& (-1)^{2J} \int^{\pi}_{-\pi} \frac{d^3 p}{(2\pi)^3}
\tr \log[1 - \kappa Q({\bf p}, J)]
\nonumber \\
&=& (-1)^{2J} \int^{\pi}_{-\pi} \frac{d^3 p}{(2\pi)^3}
\log \det [1 - \kappa Q({\bf p}, J)] .
\label{freeenergy2}
\ea
Here the second equality of the above equation is only formal one,
because the convergence of the infinite series of
the r.h.s.\ of the first line in eq.(\ref{freeenergy2})
is not ensured.
In appendix, we show that there exists a critical diffusion
constant $\kappa_c$ such that for any $|\kappa|< |\kappa_c|$
the infinite summation converges uniformly with respect to ${\bf p}$.
There it is shown $\kappa_c$ is the inverse of the eigenvalue of
the matrix $Q({\bf 0}, J)$ whose absolute value is not less than
that of other eigenvalues.
The free energy is singular at $\kappa=\kappa_c$.
As we will see, a continuum field theory is constructed
by taking the limit $\kappa \rightarrow \kappa_c$.

Therefore we have to calculate the eigenvalues of the matrix
$Q({\bf 0}, J)$, that is, to solve the equation
\be
  \det [x - Q({\bf 0}, J)] = 0 .
  \label{characteristicequation}
\ee
Note that this equation has
the following symmetries
\ba
J &\rightarrow& J + 4,    
\nonumber \\
J &\rightarrow& -J .
\ea
The former is obvious from the eq.(\ref{PROQ}).
The latter is due to the fact that the transposed matrix
of $Q({\bf 0},J)$ is equal to $Q({\bf 0}, -J)$.
Moreover, we can show that the free energy possesses these
symmetries as well.
Thus, there are only five independent models corresponding to
\be
J= \left\{ 0, \; \frac{1}{2}, \; 1, \; \frac{3}{2}, \; 2 \right\} .
\ee

By solving the algebraic equation (\ref{characteristicequation})
of six-degree in $x$, we can obtain all eigenvalues of $Q({\bf 0}, J)$.
They are listed in table \ref{tableofeigenvalues}.
The degeneracy of the eigenvalue with the largest absolute value
for each $J$ is $1$, $2$, $3$  and $4$ for $J=0,1/2,1$ and $3/2$
respectively.

\begin{table}[b]
\begin{center}
\renewcommand{\arraystretch}{1.2}
\begin{tabular}{|l||llllll|} \hline
{\it Spin Factor} & \multicolumn{6}{c|}{\it Eigenvalues} \\ \hline\hline
J=0 & {\bf 5} & 1 & 1 & 1 & $-1$ & $-1$ \\ \hline
J=1/2 & ${\bf 1+2\sqrt{2}}$ & ${\bf 1+2\sqrt{2}}$ & $1-\sqrt{2}$ &
        $1-\sqrt{2}$ & $1-\sqrt{2}$ & $1-\sqrt{2}$ \\ \hline
J=1 & {\bf 3} & {\bf 3} & {\bf 3} & $-1$ & $-1$ & $-1$ \\ \hline
J=3/2 & ${\bf 1+\sqrt{2}}$ & ${\bf 1+\sqrt{2}}$ & ${\bf 1+\sqrt{2}}$ &
        ${\bf 1+\sqrt{2}}$ & $1-2\sqrt{2}$ & $1-2\sqrt{2}$ \\ \hline
J=2 & {\bf 3} & {\bf 3} & 1 & 1 & 1 & ${\bf -3}$ \\ \hline
\end{tabular}
\end{center}
\caption[Table]{{\footnotesize Table of eigenvalues of the matrix $Q({\bf
0}, J)$. The eigenvalues with the largest absolute value
are written in boldface.}}
\label{tableofeigenvalues}
\end{table}

We  represent the explicit form of the free energy density
for each values of spin $J$.
\ba
f(\kappa, J) &=& (-1)^{2J} \int^{\pi}_{-\pi} \frac{d^3 p}{(2\pi)^3} \times
\tilde{f}({\bf p} | \kappa, J),
\nonumber \\
\tilde{f}({\bf p} | \kappa, J=0)  &=&
     \log (\kappa^2-1) (\kappa^2+1)
          (5\kappa^2 - 2\kappa\sum^{3}_{i=1} \cos p_i +1) \nonumber \\
\tilde{f}({\bf p} | \kappa, J=1/2) &=&
    \log \left[(57-40\sqrt{2}) \kappa^6 + 3\kappa^2 + 1
          +2(-5+4\sqrt{2}) \kappa^5 \sum^{3}_{i=1}\cos p_i \right.\nonumber \\
      & & +\kappa^4 \left(3+8(-1+\sqrt{2})(\cos p_1 \cos p_2 +
          \cos p_2 \cos p_3 + \cos p_3 \cos p_1)\right) \nonumber \\
      & & \left.
          +4\kappa^3 \left(\sum^{3}_{i=1}\cos p_i + 2(2-\sqrt{2})
          \prod^{3}_{i=1}\cos p_i \right)
          -2\kappa \sum^{3}_{i=1}\cos p_i
          \right] \nonumber \\
\tilde{f}({\bf p} | \kappa, J=1) &=&
     \log \left[-27\kappa^6 - 9\kappa^4 + 3\kappa^2 + 1
          \right.\nonumber \\
      & & \left.
          - 18\kappa^5 \sum^{3}_{i=1} \cos p_i
          + 4\kappa^3 \left(\sum^{3}_{i=1} \cos p_i + 4 \prod^{3}_{i=1}
          \cos p_i\right)
          \right] \\
\tilde{f}({\bf p} | \kappa, J=3/2) &=&
     \log \left[(57 + 40 \sqrt{2}) \kappa^6 + 3 \kappa^2 + 1
          - (10 + 8 \sqrt{2}) \kappa^5 \sum^{3}_{i=1}\cos p_i
          \right.\nonumber \\
      & & + \kappa^4 \left(3 - 4(1+\sqrt{2}) (\cos p_1 \cos p_2 +
          \cos p_2 \cos p_3 + \cos p_3 \cos p_1) \right) \nonumber \\
      & & \left.
          + \kappa^3 \left(4 \sum^{3}_{i=1} \cos p_i
          + 8 (2 +\sqrt{2}) \prod^{3}_{i=1}\cos p_i \right)
          -2 \kappa \sum^{3}_{i=1} \cos p_i
          \right]
          \nonumber \\
\tilde{f}({\bf p} | \kappa, J=2) &=&
     \log \left[-27 \kappa^6 + 3 \kappa^2 + 1
          + 30 \kappa^5 \sum^{3}_{i=1}\cos p_i \right.\nonumber \\
      & & - \kappa^4 \left(9 + 32(\cos p_1 \cos p_2 +
          \cos p_2 \cos p_3 + \cos p_3 \cos p_1)\right) \nonumber \\
      & & \left.
          + 4 \kappa^3 \left(\sum^{3}_{i=1}\cos p_i
          + 8 \prod^{3}_{i=1}\cos p_i \right)
          - 2 \kappa \sum^{3}_{i=1}\cos p_i
                    \right]    \nonumber
\label{explicitfreeenergy}
\ea
%

The two-point function is defined as
\be
K^{\alpha\beta}({\bf x} | \kappa, J) =
\lim_{L \rightarrow \infty}\sum_{l=0}^{L}
(Q^{-1} ({\bf 0}, J))^{\alpha\gamma} \cdot
A^{\gamma\beta}_{l}({\bf x}| \kappa,J).
\label{correlationfunction}
\ee
Here, the sum is taken over all walks which start the origin
{\it toward} the direction ${\bf e}_{\alpha}$ and come into
a site ${\bf x}$ from the direction ${\bf e}_{\beta}$.
Using eq.(\ref{fourierrecursion}) and the condition
(\ref{initialcondition}),
the two-point function (\ref{correlationfunction}) is rewritten as
\be
K^{\alpha\beta}({\bf x} | \kappa, J) =
\int^{\pi}_{-\pi} \frac{d^3 p}{(2\pi)^3}
\e^{i{\bf p}\cdot{\bf x}} \tilde{K}^{\alpha\beta}({\bf p} | \kappa, J)  ,
\label{fouriercorrelation}
\ee
where we use the following notation
$$
\tilde{K}^{\alpha\beta}({\bf p} | \kappa, J) =
\left[\left[ (1 - \kappa Q({\bf p}, J)) Q({\bf 0}, J)
\right]^{-1}\right]^{\alpha\beta} .
$$

%
%
\newsection{Critical Behavior}
In the previous section, we have defined a random walk model
with complex-valued spin factor.
In this section, we argue critical behavior of our model
near $\kappa_c$ ($|\kappa| < |\kappa_c|$).
As we have already discussed, there are only five independent models
on the cubic lattice.
They correspond to the case
$J=0$, $1/2$, $1$, $3/2$, $2$.
We discuss the critical behavior for each $J$.
Here we evaluate the singular part of the free energy
(\ref{freeenergy2}) with respect to the following variable
$$
s = \frac{\kappa_c - \kappa}{\kappa_c}  .
$$

1) $J=0$ case \\ 
In the case of $J=0$, the phase factor is trivial
as  shown in the eq.(\ref{PROQ}).
This model is well-known as Brownian motion.
The free energy density at low energy is
\be
f(\kappa, J=0) = \int^{{\bf p} \sim {\bf 0}}
\frac{d^3 p}{(2\pi)^3} \log
\left[  \frac{576}{3125} \left({\bf p}^2 + 4 s \right)
+ \cdots
\right]  .
\label{scalingfreeenergyn0}
\ee
The notation ${\bf p} \sim {\bf 0}$ denotes the expansion
around ${\bf p} = {\bf 0}$.
The most singular part of the free energy density
with respect to $s$ is
\ba
f(\kappa, J=0) \sim s^{\frac{3}{2}} \log s \quad (\kappa \nearrow 1/5).
\ea
Here we see the phase transition is of second order,
therefore the exponent  $\alpha=\frac{1}{2}$ is obtained.
This result agrees with the well-known result of Brownian motion.

2) $J=1/2$ case \\ 
The low energy limit of the free energy density  for $J=1/2$ is
\ba
f(\kappa, J=1/2) =  \int^{{\bf p} \sim {\bf 0}}
\frac{d^3 p}{(2\pi)^3} \log
\left[ \frac{36}{(1 + 2\sqrt{2})^4} \left({\bf p}^2
+ 9 s^2 \right)
+ \cdots
\right]  .
\label{scalingfreeenergyn1}
\ea
The most singular part of the free energy density is
\ba
f(\kappa, J=1/2) \sim s^3 \log s \quad (\kappa \nearrow (1+2\sqrt{2})^{-1}).
\ea
Therefore the phase transition is of third order, and the exponent
$\alpha = -1$ is obtained.

3) $J=1$ case \\ 
The critical behavior of the $J=1$ model is more complicated
than the former two cases because there are several massless excitations.
The singular part of the free energy is dominated by the  integral region
near ${\bf p} = (\pi,0,0), (0,\pi,0), (0,0,\pi)$
as well as ${\bf p} = {\bf 0}$ :
\ba
f(\kappa, J=1) &=& \int^{{\bf p} \sim {\bf 0}}
\frac{d^3 p}{(2\pi)^3} \log
\left[ \frac{16}{27} \left(s {\bf p}^2 \right.\right.
+ 4 s^3  \nonumber \\
&+& \left.\left.\frac{1}{4}
(p_1{}^2 p_2{}^2 + p_2{}^2 p_3{}^2 + p_3{}^2 p_1{}^2)  \right)
+ \cdots \right] \nonumber \\
&+& \int^{{\bf p} \sim (\pi,0,0)} \frac{d^2 p}{(2\pi)^2} \log
\left[ \frac{16}{27}(p_2{}^2 + p_3{}^2 + 4 s )
+ \cdots \right]  \nonumber \\
&+& \int^{{\bf p} \sim (0,\pi,0)} \frac{d^2 p}{(2\pi)^2} \log
\left[ \frac{16}{27} (p_1{}^2 + p_3{}^2 + 4 s )
+ \cdots \right]  \nonumber \\
&+& \int^{{\bf p} \sim (0,0,\pi)} \frac{d^2 p}{(2\pi)^2} \log
\left[ \frac{16}{27} (p_1{}^2 + p_2{}^2 + 4 s )
+ \cdots \right] . \nonumber \\
\label{scalingfreeenergyn2}
\ea
The last three excitations do {\it not} ~recover
the three-dimensional rotational invariance even near the critical point.
They are all two-dimensional scalar fields.
The most singular part in $s$ of the free energy density comes
from the last three terms in (\ref{scalingfreeenergyn2})
\ba
f(\kappa, J=1) \sim
s \log s \quad (\kappa \nearrow 1/3).
\ea
We can read from this equation that the phase transition
is of first order
\footnote{One must exercise care to determine the true order
of the phase transition for $J=1$, because the excitations near
${\bf p} = (\pi,0,0), (0,\pi,0), (0,0,\pi)$
resemble those in two dimensional Brownian motion which is ill-defined.
In practice, two dimensional Brownian motion has I.R. divergences.
}.

4) $J=3/2$ case \\ 
In this case, there are three excitations near
${\bf p}=(\pi,0,0),(0,\pi,0),(0,0,\pi)$ as well as in the $J=1$ case.
The degeneracy of the eigenvalue $1+\sqrt{2}$
of the matrix $Q({\bf p}, J=3/2)$ for ${\bf p}={\bf 0}$ is 4,
but that of the three other excitations is 2.
The low energy scheme of the free energy density for $J=3/2$ leads to
\ba
f(\kappa, J=3/2) &=&  \int^{{\bf p} \sim {\bf 0}}
\frac{d^3 p}{(2\pi)^3} \log
\left[ \frac{32}{(1 + \sqrt{2})^2} \left( s^2 ({\bf p}^2
+ \frac{9}{4} s^2) \right.\right. \nonumber \\
&+& \left.\left. 
\frac{1}{4} (p_1{}^2 p_2{}^2 + p_2{}^2 p_3{}^2 + p_3{}^2 p_1{}^2) \right)
+ \cdots \right]  \nonumber \\
&+& \int^{{\bf p} \sim (\pi,0,0)} \frac{d^2 p}{(2\pi)^2} \log
\left[ \frac{32}{(1 + \sqrt{2})^2} (p_2{}^2 + p_3{}^2+ 4 s^2)
+ \cdots \right]  \nonumber \\
&+& \int^{{\bf p} \sim (0,\pi,0)} \frac{d^2 p}{(2\pi)^2} \log
\left[ \frac{32}{(1 + \sqrt{2})^2} (p_1{}^2 + p_3{}^2 + 4 s^2)
+ \cdots \right]  \nonumber \\
&+& \int^{{\bf p} \sim (0,0,\pi)} \frac{d^2 p}{(2\pi)^2} \log
\left[ \frac{32}{(1 + \sqrt{2})^2} (p_1{}^2 + p_2{}^2 + 4 s^2)
+ \cdots \right] . \nonumber \\
\label{scalingfreeenergyn3}
\ea
Note that the term
$p_1{}^2 p_2{}^2 + p_2{}^2 p_3{}^2 + p_3{}^2 p_1{}^2$
breaks rotational invariance.
The most singular part with respect to  $s$ is
\be
f(\kappa, J = 3/2) \sim s^{2} \log s \quad
(\kappa \nearrow (1+\sqrt{2})^{-1}).
\ee
The phase transition is of second order, and the exponent $\alpha$
is $0$.

5) $J=2$ case \\ 
There are two critical points
$\kappa_c = 1/3$ and $\kappa_c = -1/3$ in this case.
In the case of $\kappa_c = 1/3$,
the low energy limit of the free energy density is
\ba
f(\kappa \sim 1/3, J=2) &=& \int^{{\bf p} \sim {\bf 0}}
\frac{d^3 p}{(2\pi)^3} \log
\left[ \frac{32}{81} \left(s({\bf p}^2
+ \frac{3}{2} s) \right.\right. \nonumber \\
&+& \left.\left. 
\frac{1}{2} (p_1{}^2 p_2{}^2 + p_2{}^2 p_3{}^2 + p_3{}^2 p_1{}^2) \right)
+ \cdots \right] .
\ea
There exists a non-rotational invariant term.
The leading behavior of the free energy density in terms of $s$ is
\be
f(\kappa, J=2)
\sim s^{\frac{3}{2}} \log s \quad (\kappa \nearrow 1/3) .
\ee
The phase transition is of second order,
and the exponent $\alpha$ is $1/2$.

In the  case  of $\kappa= -1/3$, the low energy limit of the free energy
density is
\be
f(\kappa\sim -1/3, J=2) = \int^{{\bf p} \sim {\bf 0}}
\frac{d^3 p}{(2\pi)^3} \log
\left[  \frac{64}{81} \left({\bf p}^2 + 12 s \right)
+ \cdots \right] .
\ee
The leading behavior of the free energy density in terms of $s$ is
\be
f(\kappa, J=2)
\sim s^{\frac{3}{2}} \log s \quad (\kappa \searrow -1/3) .
\ee
The phase transition is of second order,
and the exponent $\alpha$ is $1/2$.

%
%
\newsection{The Scaling Limit}
In this section, we derive the $n$-point correlation functions
for continuum spinor field theories.

We restore the lattice spacing parameter $a$ by the following replacement :
\ba
{\bf x} &\rightarrow& {\bf x}/a \nonumber \\
{\bf p} &\rightarrow& {\bf p}\cdot a \\
s &=& (m \cdot a)^{\frac{1}{\nu}} . \nonumber
\label{scaling}
\ea
Here the exponent $\frac{1}{\nu}$ must be chosen in such a way
that the correlation functions have suitable continuum limit,
the scaling limit.
This can be achieved by taking $a \rightarrow 0$, $s \rightarrow 0$ while
keeping $m$ finite.
We investigate the scaling limit of the correlation functions.
We have to evaluate only the two-point function since
all the theories we construct are free field theories.
We can see whether they have rotational invariance or not.

To extract the massless mode from the two-point function
$K^{\alpha\beta}({\bf x}|\kappa,J)$, we choose a unitary matrix
$U({\bf 0}, J)$ which diagonalizes the mass matrix
$1 - \kappa Q({\bf 0}, J)$.
For this purpose, we introduce
\be
K_{{\rm diag}}^{\alpha\beta}({\bf x}|\kappa, J) \stackrel{{\rm def}}{=}
U({\bf 0}, J) \cdot K^{\alpha\beta}({\bf x}|\kappa, J) \cdot
U({\bf 0}, J)^{\dagger} .
\ee
We choose $\nu = 1/2,\ 1,\ 1$ for $J=0,\ 1/2,\ 1$ respectively
for the scaling limit.
The dominant part of the two-point correlation functions
$K^{\alpha\beta}_{diag}$ for $J=0,\ 1/2,\ 1$ becomes
\ba
K_{{\rm diag}}^{\alpha\beta}({\bf x}|\kappa, J=0) &=&
\frac{4 a}{5} \int^{\pi/a}_{-\pi/a} \frac{d^3 p}{(2\pi)^3}
\e^{i{\bf p}\cdot{\bf x}}
\left(
\renewcommand{\arraystretch}{2}
\begin{array}{c|c}
\frac{\mbox{\large 1}}{\mbox{\large \bf p}^2 \mbox{\large + 4 m}^2}
& \mbox{{\large 0}}_{(1\;5)} \\
\hline
\mbox{{\large 0}}_{(5\;1)} & \mbox{{\large 0}}_{(5\;5)} \\
\end{array}
\right)
+ O(a^2)
\nonumber \\
K_{{\rm diag}}^{\alpha\beta}({\bf x}|\kappa, J=1/2) &=&
\frac{3 a^2}{1 + 2\sqrt{2}} \int^{\pi/a}_{-\pi/a} \frac{d^3 p}{(2\pi)^3}
\e^{i{\bf p}\cdot{\bf x}}
\left(
\renewcommand{\arraystretch}{2}
\begin{array}{c|c}
\frac{\mbox{\large 1}}
{\mbox{{\large i}} \mbox{\large{\boldmath
$\sigma$}} \cdot
\mbox{{\large {\bf p}}} \; \mbox{{\large + 3 m I}}_{(2\;2)}}
& \mbox{{\large 0}}_{(2\;4)} \\
\hline
\mbox{{\large 0}}_{(4\;2)} & \mbox{{\large 0}}_{(4\;4)} \\
\end{array}
\right)   \nonumber \\
& & + O(a^3)
\nonumber \\
K_{{\rm diag}}^{\alpha\beta}({\bf x}|\kappa, J=1) &=&
\frac{2 a^2}{3} \int^{\pi/a}_{-\pi/a} \frac{d^3 p}{(2\pi)^3}
\e^{i{\bf p}\cdot{\bf x}} \nonumber \\
& & \left[ \left(
\renewcommand{\arraystretch}{2}
\begin{array}{c|c}
\frac{\mbox{{\large 1}}}
{\mbox{{\large i}} \mbox{{\large \bf L}} \cdot
\mbox{{\large \bf p}} \;
\mbox{{\large + 2 m I}}_{(3\;3)}} &
\mbox{{\Large 0}}_{(3\;3)} \\
\hline
\mbox{{\Large 0}}_{(3\;3)} & \mbox{{\Large 0}}_{(3\;3)} \\
\end{array}
\right)
+ \left( \begin{array}{c}
       {\rm {\bf p} \ indep.}  \\
       {\rm matrix}
       \end{array}
\right)
\right]  \nonumber \\
& & + O(a^3) .
\ea
Here $0_{(i\;j)}$ denotes an $i\times j$ zero matrix
and  $I_{(i\;i)}$ is the $i \times i$ identity matrix.
The matrices $\mbox{\large{\boldmath $\sigma$}}$ are the Pauli matrices
and the $3 \times 3$ matrices ${\bf L}$ are the spin-1 matrices defined by
$[L_i,L_j] = i \epsilon_{ijk} L_k$ and ${\bf L}\cdot{\bf L}=2$.
The ${\bf p}$ independent matrix appears in
$ K_{{\rm diag}}({\bf p}| \kappa, J=1)$
is originated from the three other excitations which do not produce
long-range correlation in the scaling limit.
Actually, the ${\bf p}$ independent part gives a $\delta$-function
in coordinate space.
We can subtract such a short range function
from the two-point correlation function.
Note that the rotational symmetry breaking terms
in the two-point functions corresponding to
those in eq.(\ref{scalingfreeenergyn2})
disappear in the scaling limit.

In the $J=3/2$ case, however, the situation is completely different
from $J=1$.
We can {\it not} obtain the three-dimensional rotational invariant
theory in the scaling limit.
Actually, whatever the choice of $\nu$ we cannot remove the breaking
term $p_1{}^2 p_2{}^2 + p_2{}^2 p_3{}^2 + p_3{}^2 p_1{}^2$ contrary
to the $J=1$ case.

The correlation functions for long range modes are
\ba
g({\bf x}| J=0) &\stackrel{\rm def}{=}&
\lim_{\stackrel{a \rightarrow 0}{s=m^2\cdot a^2}} \frac{5}{4 a} \cdot
K_{{\rm diag}}^{\alpha\beta}({\bf x}|\kappa, J=0)
\nonumber \\
&=&
\int^{\infty}_{-\infty} \frac{d^3 p}{(2\pi)^3}
\e^{i{\bf p}\cdot{\bf x}}
\left(
\renewcommand{\arraystretch}{2}
\begin{array}{c|c}
\frac{\mbox{\large 1}}{\mbox{\large \bf p}^2 \mbox{\large + 4
    m}^2}
& \mbox{{\large 0}}_{(1\;5)} \\
\hline
\mbox{{\large 0}}_{(5\;1)} & \mbox{{\large 0}}_{(5\;5)} \\
\end{array}
\right) \nonumber \\
g({\bf x}| J=1/2) &\stackrel{\rm def}{=}&
\lim_{\stackrel{a \rightarrow 0}{s=m\cdot a}}
\frac{1 + 2\sqrt{2}}{3 a^2} \cdot
K_{{\rm diag}}^{\alpha\beta}({\bf x}|\kappa, J=1/2)
\nonumber \\
&=&
\int^{\infty}_{-\infty} \frac{d^3 p}{(2\pi)^3}
\e^{i{\bf p}\cdot{\bf x}}
\left(
\renewcommand{\arraystretch}{2}
\begin{array}{c|c}
\frac{\mbox{\large 1}}
{\mbox{{\large i}} \mbox{\large{\boldmath
$\sigma$}} \cdot
\mbox{{\large {\bf p}}} \; \mbox{{\large + 3 m I}}_{(2\;2)}}
& \mbox{{\large 0}}_{(2\;4)} \\
\hline
\mbox{{\large 0}}_{(4\;2)} & \mbox{{\large 0}}_{(4\;4)} \\
\end{array}
\right) \nonumber \\
g({\bf x}| J=1) &\stackrel{\rm def}{=}&
\lim_{\stackrel{a \rightarrow 0}{s=m\cdot a}}
\frac{3}{2 a^2} \cdot K_{{\rm diag}}^{\alpha\beta}({\bf x}|\kappa, J=1)
\nonumber \\
&=&
\int^{\infty}_{-\infty} \frac{d^3 p}{(2\pi)^3}
\e^{i{\bf p}\cdot{\bf x}}
\left(
\renewcommand{\arraystretch}{2}
\begin{array}{c|c}
\frac{\mbox{{\large 1}}}
{\mbox{{\large i}} \mbox{{\large \bf L}} \cdot
\mbox{{\large \bf p}} \;
\mbox{{\large + 2 m I}}_{(3\;3)}} &
\mbox{{\Large 0}}_{(3\;3)} \\
\hline
\mbox{{\Large 0}}_{(3\;3)} & \mbox{{\Large 0}}_{(3\;3)} \\
\end{array}
\right)  ,
\ea
where we removed the $\delta$-functions of ${\bf x}$ in the $J=1$ case.
We have  constructed the two-point correlation functions
for the Klein-Gordon, Dirac and the $U(1)$ Chern-Simons fields
from our random walk model with $J=0,\ 1/2$ and $1$ respectively.
Therefore these free field theories are obtained rigorously.
On the other hand, the spinor theories with spin higher than
$1$ cannot be obtained from this construction.

Now we discuss critical exponents and relations among them.
The susceptibility matrix can be defined
by analogy with classical spin systems
\ba
\chi^{\alpha\beta}_{{\rm diag}}(\kappa, J) &=&
\sum_{{\bf x}} K^{\alpha\beta}_{{\rm diag}}({\bf x}|\kappa, J)
\nonumber \\
&=& \left( \frac{1}{(1 - \kappa Q({\bf 0}, J))
\cdot Q({\bf 0}, J)}\right)^{\alpha\beta}_{{\rm diag}}
\label{susce}
\ea
The susceptibility exponent (matrix) $\mbox{\large{$\gamma$}}$ is
defined from the behavior of $\chi$ near the critical point in $s$
\be
\chi^{\alpha\beta}_{{\rm diag}}
\sim (\kappa_c - \kappa)^{-\mbox{\large{$\gamma$}}^{\alpha\beta}}
\sim s^{-\mbox{\large{$\gamma$}}^{\alpha\beta}}.
\ee
We see the matrix $\mbox{\large{$\gamma$}}^{\alpha\beta}$
is the identity matrix of order $1$, $2$ and $3$
for $J=0,1/2$ and $1$ respectively.

The power behaviors in $|{\bf x}|$ at long-distance of
the correlation functions are
\ba
\lim_{|{\bf x}| \rightarrow \infty}
g({\bf x}|J=0) & \sim &
  \frac{1}{|{\bf x}|} \nonumber \\
\lim_{|{\bf x}| \rightarrow \infty}
g({\bf x}|J=1/2) & \sim &
  \frac{\mbox{\large{\boldmath $\sigma$}} \cdot
  {\bf e}_{({\bf x})}}{|{\bf x}|^2}
\nonumber \\
\lim_{|{\bf x}| \rightarrow \infty}
g({\bf x}|J=1) &\sim &
  \frac{{\bf L} \cdot {\bf e}_{({\bf x})}}{|{\bf x}|^2} ,
\label{}
\ea
where ${\bf e}_{({\bf x})}
=(x_1/|{\bf x}|, x_2/|{\bf x}|, x_3/|{\bf x}|)$.
Thus the exponent $\eta$ for each $J$ is
$\eta = 0, \ 1, \ 1$ for $J = 0,\ 1/2,\ 1$ respectively.

The exponential behavior at large-distance is also important.
The correlation length is the inverse of the mass-matrix
\be
\xi^{\alpha\beta}_{{\rm diag}}
= \left( \left( \frac{1}{(1 - \kappa Q({\bf 0}, J))
\cdot Q({\bf 0}, J)} \right)^{\nu} \right)^{\alpha\beta}_{{\rm diag}}
\ee
where $\nu$ is that of $s=(m \cdot a)^{\frac{1}{\nu}}$.
The exponent (matrix) $\mbox{\large{$\nu$}}^{\alpha\beta}$ is defined as
\be
\xi^{\alpha\beta}_{{\rm diag}} \sim s^{- \mbox{\large{$\nu$}}^{\alpha\beta}} .
\ee
In general, $\mbox{\large{$\nu$}}$ is also a matrix-valued exponent.
We can easily obtain $\mbox{\large{$\nu$}} = 1/2, I_{(22)}, I_{(33)}$
for $J=0,1/2,1$ respectively.
Here $I_{(ii)}$ is the $i \times i$ identity matrix.

We summarize our results for the critical exponents
in the table \ref{tableofexponents}.
There we show the eigenvalues of the exponent matrices
$\mbox{\large{$\nu$}}^{\alpha\beta}$ and
$\mbox{\large{$\gamma$}}^{\alpha\beta}$, $\nu$ and $\gamma$
respectively.
%
\begin{table}[hbtp]
\begin{center}
\renewcommand{\arraystretch}{1.2}
\newcommand{\lw}[1]{\smash{\lower2.ex\hbox{#1}}}
\begin{tabular}{|l||c|c|c|c|c|ll|}
\hline
{\it Spin Factor} & {\it Continuum Field Theory} & $\alpha$ & $\gamma$ &
$\eta$ & $\nu$ & $\kappa_c{}^{-1}$ &  \\
\hline\hline
J=0 & {\rm free scalar field theory} & $1/2$ & $1$ & $0$ & $1/2$ &
$5$ & (1) \\
\hline
J=1/2 & {\rm Dirac field theory} & $-1$ & $1$ & $1$ & $1$ &
$1 + 2\sqrt{2}$ & (2) \\
\hline
J=1 & {\rm massive $U(1)$ CS field theory} & $(1)$ & $1$ & $1$ &
$1$ & $3$ & (3) \\
\hline
J=3/2 &
\setlength{\unitlength}{1mm}
\begin{picture}(30,1)
\put(0,1){\line(1,0){30}}
\end{picture}
& $0$ & $-$ & $-$ & $-$ & $1 + \sqrt{2}$ & (4) \\
\hline
\lw{J=2} &
\setlength{\unitlength}{1mm}
\begin{picture}(30,1)
\put(0,1){\line(1,0){30}}
\end{picture}
& $1/2$ & $-$ & $-$ & $-$ & $3$ & (2) \\
\cline{2-8}
 & free scalar field theory & $1/2$ & $1$ & $0$ & $1/2$ & $-3$ & (1) \\
\hline
\end{tabular}
\end{center}
\caption[table]{{\footnotesize Field theories and Critical exponents
are obtained
as scaling limit of correlation functions.
The exponents $\mbox{\large{$\gamma$}}$ and $\mbox{\large{$\nu$}}$ are
generally matrix-valued, and here we show only eigenvalues.
The exponent $\alpha$ for $J=1$ must suffer I.R. divergences.
The Fisher's scaling law $\gamma = \nu (2 - \eta)$ is satisfied.
However, the hyper-scaling relation is broken for $J=1$,
rather inequality $\alpha > 2 - d \nu, (d=3)$ is satisfied.
 }}
\label{tableofexponents}
\end{table}

%
%

%
%
\newsection{Studies on Other Lattices}
So far we have used only the cubic lattice to define spinor fields on it.
With another lattice structure however, one might obtain
different results from those in the cubic lattice.
Thus if we want to claim with certainty that we obtain fields of
spin $0$, $\frac 1 2$ and $1$ from random walk models in three-dimensions,
we have to check the universality of our results by taking other
lattices as well.

In this section, we mainly consider two different lattices,
on which to define a random walk theory,
a triclinic lattice and a body-centered lattice.
We investigate the theory on these lattices and compare the results
to those on the cubic lattice.
We show that the spin $0$ and $1/2$ fields are constructed from
any lattice we study here.
On the other hand, we find an example of lattice
on which the spin $1$ field is never obtained.

\paragraph{Definition}
First, we consider a natural definition of a random walk theory on
an arbitrary periodic lattice $L$ in three dimensional space $R^3$.
A lattice $L$ is defined by putting a set of sites in $R^3$
periodically as a crystal with bonds which connect
pairs of nearest neighbor sites.
One must be concerned for the definition of the spin factor
on an arbitrary lattice.

In the case of the cubic lattice, the spin factor is determined by
the oriented area on $S^2$ encircled by unit tangent vectors of
a closed path.
We see from the diagram in Fig.\ref{asanozu1}
that the phase factor is defined as the area of the region
on $S^2$ whose boundary is determined by points of the unit tangent
vectors of a closed walk.
The area of $S^2$ is normalized as $4 \pi$.
In this way, we obtain phase factor $S(\beta, \gamma)$ of a path
attached when the path turns from a direction $\beta$ to
a direction $\gamma$ by reading Fig.\ref{asanozu1} directly.
Note that there is an ambiguity in choosing  $S(\beta,\gamma)$
which give the same oriented area.

On the basis of this spin factor, we define a random walk theory
on an arbitrary lattice $L$.
We assume non-backtracking walks as before.
We define the amplitude of walks by the recursion relation
  \begin{equation}
      A^{\alpha\beta}_{l+1}({\bf x}) = \sum_{\gamma \neq -\beta}
     \kappa^{|{\bf e}_\beta|} {\rm e}^{iJS(\beta,\gamma)}
      A^{\alpha\gamma}_{l}({\bf x}-{\bf e}_{\beta})
  \end{equation}
with the initial condition (\ref{initialcondition}).
Here $|{\bf e}_\beta|$ is the length of the bond ${\bf e}_\beta$
which is between $l$-th site and $(l+1)$-th site of the walk.
$S(\beta,\gamma)$ is the area on $S^2$ as described above.
We define the free energy density with the help of $A^{\alpha\beta}$
as in (\ref{freeenergy}).

This definition of the free energy is applicable to any periodic lattice.
Thus one might expect that spinor fields can be obtained from any lattice
as well as the regular cubic lattice.

In order to compute the free energy of the theory explicitly, we must
require two additional properties on the lattice $L$:
the length of all bonds is the same, and all sites of the lattice $L$
are equivalent.

Under these requirements, the free energy density on the lattice $L$
can be written using the Fourier transformed amplitude
$\tilde A^{\alpha\beta}_{l}({\bf p})$ and the recursion relation
corresponding to (\ref{fourierrecursion}) :
  \begin{eqnarray}
        f_L(\kappa, J) & =&  -(-1)^{2J} 
        \sum_{l=1}^{\infty} \sum_{\alpha} \frac{1}{l}
        A^{\alpha\alpha}_{l}({\bf 0})
    \nonumber \\
    &=& (-1)^{2J} \int^{\pi}_{-\pi} \frac{d^3 p}{(2\pi)^3}
      \log \det [1 - \kappa Q_L({\bf p}, J)] .
  \end{eqnarray}
Here $Q_L({\bf p},J)\equiv Q_L({\bf 0}, J) \cdot P_L({\bf p})$
and $\alpha$ denotes the direction of a bond coming out of each site
in the lattice.

If we choose a lattice $L$, i.e., determine a relation between
sites and bonds in $L$, we can fix a unit tangent vector
${\bf e}_{\alpha}$ directed to a bond $\alpha$ $(\alpha = 1,2,...,2n)$.
Here  $2n$ is the number of bonds come out of one site on $L$.
We assume ${\bf e}_{\alpha+n} = -{\bf e}_\alpha$.
Now we obtain the matrix $P_L({\bf p})$ by an analogy with (\ref{PROQ}) :
  \begin{equation}
    P_L({\bf p}) =
     \left(
        \matrix{
        e^{i p_1 '} & & & & & \bigzerou &  &\cr
         & e^{i p_2 '} & & & & & &\cr
         & & \cdot & & & & &\cr
         & & & e^{i p_n '} & & & &\cr
         & & & & e^{-i p_1 '}  & & &\cr
         & & & & & \cdot & & \cr
         & & & & & & \cdot & \cr
         & \bigzerol & & & & & & e^{-i p_n '} \cr
        }
        \right) ,
        \label{plp}
  \end{equation}
where the value $p_{\alpha} '$ is defined
as an inner product of ${\bf p}$ and ${\bf e}_\alpha$, i.e.,
 \begin{eqnarray}
   p_{\alpha} ' & \equiv & {\bf p} \cdot {\bf e}_{\alpha}
    \nonumber \\
        & = & (p_x, p_y, p_z)\cdot {\bf e}_{\alpha}
    \label{defofpalpha}
     \end{eqnarray}
The matrix $Q_L({\bf 0},J)$ is determined by depicting  a `diagram
of the phase assignment' as in Fig.\ref{asanozu1}.

Note that the matrix $Q({\bf p},J)$ and the free energy constructed
from it have the symmetry $J \rightarrow -J$ as in the case of
cubic lattice, i.e., $Q_L({\bf p}, -J) = Q_L{}^{t}({\bf p}, J)$.
However, the symmetry $J \rightarrow J+4$ does not hold any more
in general.
It follows that the possibility of higher spin fields appearing now
exists.

In the case of $J=0$, the explicit form of $Q({\bf 0},J=0)$ is
\begin{equation}
      Q({\bf 0}, J=0) = \left(
\matrix{1 & \cdot & \cdot &\cdot & 1 & 0 & 1 &\cdot &\cdot &1 \cr
  \cdot & \cdot & & & \cdot & 1 & \cdot & & &\cdot \cr
   \cdot & & \cdot & & \cdot & \cdot & & \cdot & &\cdot \cr
   \cdot & & & \cdot &\cdot &\cdot & & & \cdot & 1 \cr
   1 & \cdot & \cdot & \cdot & 1 & 1 & \cdot & \cdot & 1 & 0 \cr
      0 & 1  & \cdot &\cdot & 1 & 1 & \cdot &\cdot &\cdot &1 \cr
  1 & \cdot & & & \cdot & \cdot & \cdot & & &\cdot \cr
   \cdot & & \cdot & & \cdot & \cdot & & \cdot & &\cdot \cr
   \cdot & & & \cdot &1 &\cdot & & & \cdot & \cdot \cr
   1 & \cdot & \cdot & 1 & 0 & 1 & \cdot & \cdot & \cdot & 1
     }       \right) .
  \label{ql0}
  \end{equation}
Some $0$ valued elements in the matrix represent the non-backtracking
property of walks.
All eigenvalues of this matrix can be shown to be either $2 n-1$,
$1$ or $-1$.
The degeneracy being $1$, $n$ and $n-1$, respectively.

The summation over all possible walks converges if and only if
$|\kappa| < \kappa_c = (2 n-1)^{-1}$.
Since the eigenvalue $2 n-1$ has no degeneracy, we expect
to obtain the Klein-Gordon theory near this critical point
as in the case of cubic lattice.
The corresponding free energy is given by
  \begin{equation}
      f(\kappa, J=0)=  \int^{\pi}_{-\pi} \frac{d^3 p}{(2\pi)^3}
      \log \det [1 - \frac{1-s}{2 n -1} Q_L({\bf p}, J=0)] .
    \label{freeunij0}
  \end{equation}

The singular part of the free energy (\ref{freeunij0}) is
\be
  f_L(\kappa, J=0) = \int^{{\bf p} \sim {\bf 0}}
  \frac{d^3 p}{(2\pi)^3} \log
  \left[  A \left( ({p_1 '}^2+{p_2 '}^2 + \cdots +{p_n '}^2) + 2(n-1) s\right)
  + \cdots
  \right] ,
\ee
where
$$
  A = \frac{2^{2n-2}(n-1)^{n-1}n^{n-1}} {(2n-1)^{2n-1}} .
$$
We see the phase transition is of second order as in the case of
cubic lattice.

For other values of $J$, it is difficult to determine the critical behavior
of the theory without specifying a lattice structure
because the form of the matrix $Q({\bf p},J)$ depends on the lattice.

In the following, we take two different lattices, triclinic lattice
and body centered lattice, and examine the theory in detail
especially in the case of $J=1/2$ and $1$.

\begin{figure}
    \begin{center}
       \leavevmode
       \epsfxsize = 4.5cm
         \epsfbox{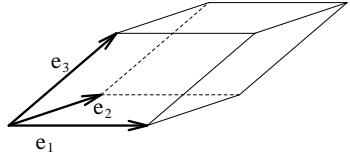}
    \end{center}
     \caption{The triclinic lattice.}
    \label{tricliniclattice}
\end{figure}

\paragraph{The Triclinic Lattice}
Here we consider the theory on a triclinic lattice, that is
a lattice defined by inclining a cubic lattice as depicted in
Fig.\ref{tricliniclattice}.

\begin{figure}
   \begin{center}
    \leavevmode
    \epsfxsize = 7cm
     \epsfbox{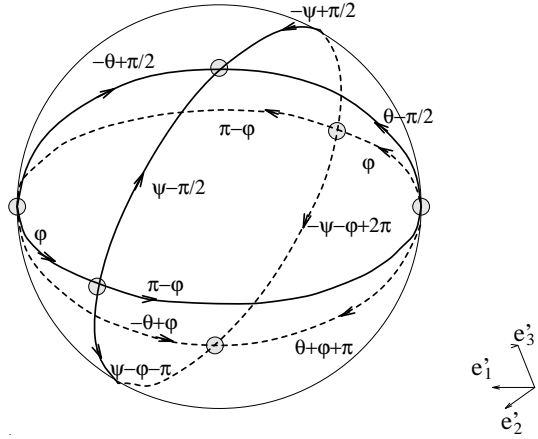}
   \end{center}
\caption{Definition of the phase of triclinic lattice.}
\label{diagramtri}
\end{figure}

If we assume that the three bonds have the same length $1$,
it is characterized by three parameters $\theta$, $\phi$ and $\psi$ each
of which represents the angle between two adjacent faces.
We call such a lattice $L_{\theta,\phi,\psi}$.
The range of these three parameters $\theta$, $\phi$ and $\psi$ is
\begin{equation}
     0 < \theta, \phi, \psi < \pi \quad , \quad
     \pi  < \theta + \phi + \psi < 3 \pi .
\end{equation}
Let us define the propagation matrix
$Q_{\theta,\phi,\psi}({\bf p},J)$ for the lattice $L_{\theta,\phi,\psi}$.
The diagram of the phase assignment is given by Fig.\ref{diagramtri},
in which the area of $S^2$ is normalized to be $4\pi$.
We can read from the diagram each element of the matrix :
\begin{equation}
Q_{\theta, \phi, \psi}({\bf 0}, J) = \left(
\matrix{
1 & e^{\phi iJ} & e^{\frac{\pi-2 \theta} 2 iJ} & 0
& e^{(\phi-\pi) iJ} & e^{(\phi-\theta) iJ } \cr
e^{-\phi iJ} & 1 & e^{\frac{2 \psi - \pi}2 iJ}
& e^{(\pi-\phi) iJ} & 0 & e^{(\psi-\phi-\pi) iJ} \cr
e^{\frac{2 \theta -\pi}2 iJ} & e^{\frac{\pi- 2\psi}2 iJ} & 1
& e^{\frac{\pi-2 \theta}2 iJ} & e^{\frac{2 \psi -\pi}2 iJ} & 0 \cr
0 & e^{(\phi-\pi) iJ} & e^{\frac{2 \theta-\pi}2 iJ}
& 1 & e^{\phi iJ} & e^{(\theta +\phi +\pi) iJ} \cr
e^{(\pi -\phi)iJ} & 0 & e^{\frac{\pi-2 \psi}2 iJ}
& e^{-\phi iJ} & 1 & e^{(2\pi-\psi-\phi) iJ} \cr
e^{(\theta-\phi) iJ} & e^{(\pi+\phi-\psi)iJ} & 0
& e^{-(\theta+\phi+\pi) iJ} & e^{(\phi+\psi-2 \pi)iJ} & 1 \cr}
\right) .
\end{equation}
On the other hand, the matrix $P({\bf p})$ is defined by unit
tangent vectors of three bond axes ${\bf e}'_1$, ${\bf e}'_2$ and
${\bf e}'_3$ of the lattice
as
  \begin{equation}
     P({\bf p}) = \left(
        \matrix{
        e^{i p_1 '} & & & & \bigzerou & \cr
         & e^{i p_2 '} & & & & \cr
         & & e^{i p_3 '} & & & \cr
         & & & e^{-i p_1 '} & & \cr
         & & & & e^{-i p_2 '} & \cr
         & \bigzerol & & & & e^{-i p_3 '} \cr
        }
        \right)
   \end{equation}
where $p_{\alpha}'\equiv {\bf p}\cdot{\bf e}'_{\alpha}$.

To investigate the scaling limit of the theory
we must find the degeneracy of the
eigenvalue $\lambda_m$ of $Q_{\theta,\phi,\psi}({\bf p},J)$
whose absolute value is not less than that of the other eigenvalues.
First, we consider the $J=1/2$ model.
Solving the eigenvalue equation explicitly, we find that
the six eigenvalues of the matrix
$Q_{\theta, \phi, \psi}({\bf 0}, J=1/2)$
always appear in degenerated pairs.
We write them as $\lambda_1$, $\lambda_1$, $\lambda_2$, $\lambda_2$,
$\lambda_3$ and $\lambda_3$.
They are continuous functions of $(\theta, \phi,\psi)$ :
$$\lambda_i=\lambda_i(\theta,\phi,\psi).$$
If we set
$$
\lambda_1 \left( \pi/2, \pi/2, \pi/2 \right) >
\lambda_j \left(\pi/2, \pi/2, \pi/2 \right) \quad (j=2,3),
$$
we have the relations $|\lambda_1| > |\lambda_2|$ and
$|\lambda_1| > |\lambda_3|$
for any values of $\theta$, $\phi$ and $\psi$.
We can show this by numerical calculation.
It follows that the largest absolute value of eigenvalues of
$Q_{\theta,\phi,\psi}({\bf 0},1/2)$ is $\lambda_1$ and its degeneracy
is 2.
We can define the free energy of the theory when
the diffusion constant is in the region
$| \kappa| < \kappa_c = \lambda_1{}^{-1}$.

The free energy can be expanded
in $s$
  \begin{equation}
     \tilde{f}_{\theta,\phi,\psi}({\bf p}|\kappa,J= 1/ 2)
       = \log \left[ A_{ij} p_i p_j + B s^2
           + O(p^4, s^3) \right] ,
  \end{equation}
where $A_{ij}$ ($i,j=1,2,3$) and $B$
are real functions of $(\theta, \phi, \psi)$.
The free energy
$\tilde{f}_{\theta,\phi,\psi}({\bf p}|\kappa,J = 1/2)$
 does not have rotational invariance, rather it has the symmetry of
an ellipsoid.
This elliptic symmetry is natural
in the sense that the original random walk
on the lattice $L_{\theta,\phi,\psi}$ does not have
cubic but triclinic lattice symmetry.
Thus to define a rotational invariant field theory,
we have to employ a new coordinate system
which restores the rotational invariance of the theory in the
scaling limit.
In this way the free energy is rewritten in a rotational invariant form
  \begin{equation}
     \tilde{f}_{\theta,\phi,\psi}({\bf p}|\kappa,J = 1/2)
        = \log [A {\bf p}^2+ B s^2+  O(p^4) ] .
  \end{equation}
This is the same form as  the cubic lattice
case (c.f. (\ref{scalingfreeenergyn1})).
Thus, we conclude that we can obtain
the Dirac spinor field from random walk theory also on
a lattice $L_{\theta,\phi,\psi}$
by taking
the scaling limit of the theory.

\medskip

Next, we consider  the $J=1$ case.
In this case, we expect that a vector field is constructed on
the lattice $L_{\theta,\phi,\psi}$ as well as on the cubic lattice.
In order to obtain this result, it is necessary to show
the triple degeneracy of eigenvalues of $Q({\bf p})$
whose absolute value is larger than that of other eigenvalues.

Explicit calculation of the eigenvalues of
$Q_{\theta,\phi,\psi}({\bf 0}, J=1)$ however shows that
there is no such eigenvalue for any value of
$\theta$, $\phi$ and $\psi$ ohter than
$\theta = \phi = \psi = \pi/2$.

It follows that in this case we cannot obtain a vector field
on the lattice $L_{\theta,\phi,\psi}$ in general.
Therefore, we conclude 
that the construction of the vector field from this random walk model
is not universal.

\begin{figure}
\begin{center}
\leavevmode
    \epsfxsize = 9cm
    \epsfbox{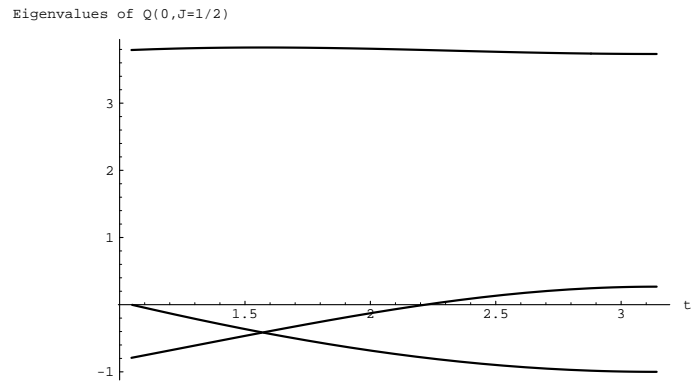}
\end{center}
\caption{Eigenvalues of $Q_t({\bf 0},J=\frac 1 2)$ }
\label{eigentj12}
\end{figure}

\begin{figure}
\begin{center}
\leavevmode
    \epsfxsize = 8cm
    \epsfbox{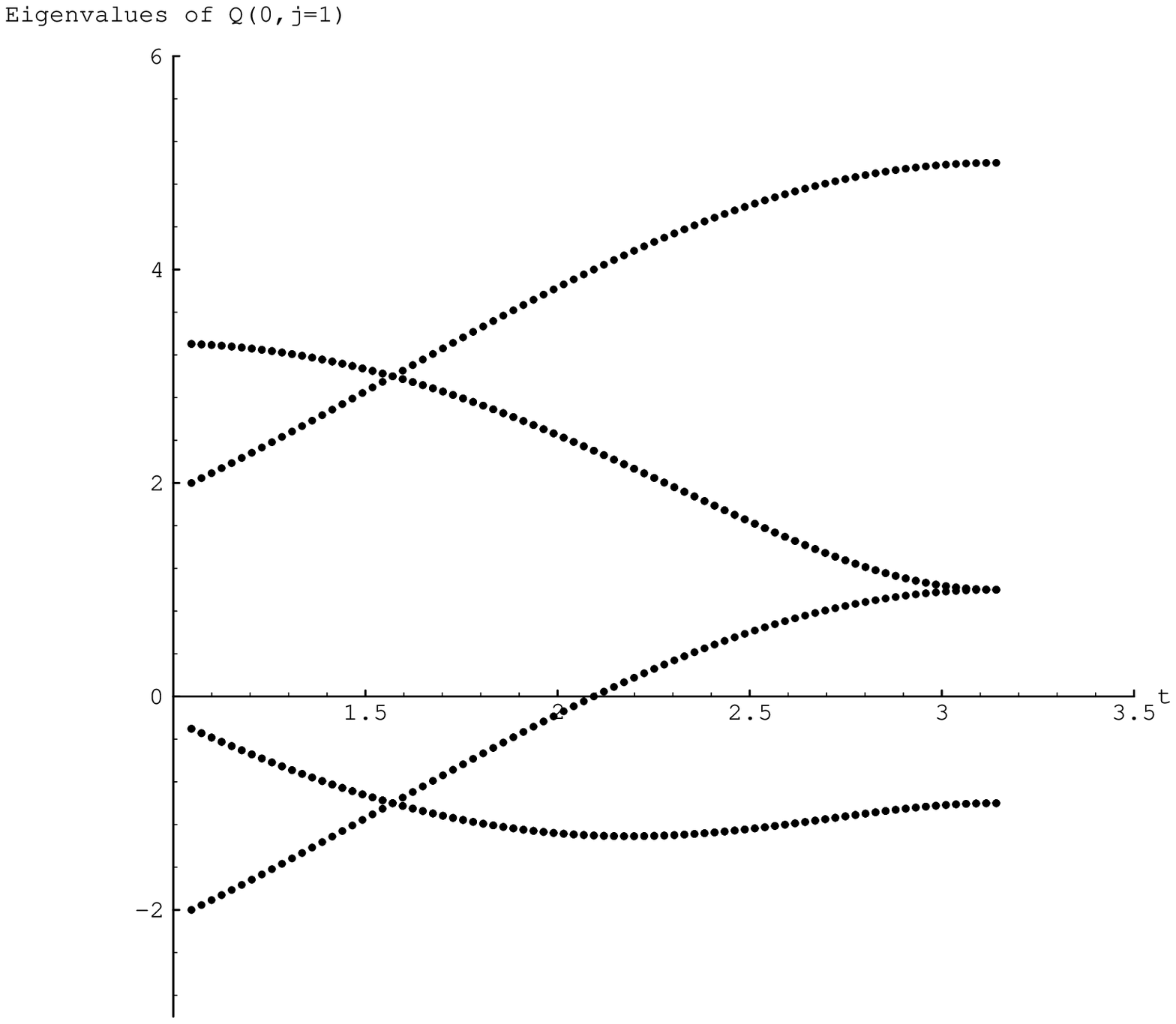}
\end{center}
\caption{Eigenvalues of $Q_t({\bf 0},J=1)$}
\label{eigentj1}
\end{figure}

\begin{figure}
\begin{center}
    \leavevmode
    \epsfysize = 3cm
     \epsfbox{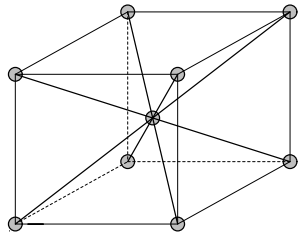}
\end{center}
\caption{Body centered lattice.}
\label{bodycenterl}
\end{figure}

To see the properties  of eigenvalues of
$Q_{\theta,\phi,\psi}({\bf 0}, J)$ in the case of $J = 1/2$ and
$J=1$, we restrict the values of parameters in
$Q_{\theta,\phi,\psi}({\bf 0}, J)$ as
$\theta =\phi =\psi (= t)$ and plot the eigenvalues of
$Q_t({\bf 0}, J)$  (Fig.\ref{eigentj12}, \ref{eigentj1}).
%
\paragraph{Body Centered Lattice}
We consider another lattice, body centered cubic lattice.
A part of the lattice is depicted in Fig.\ref{bodycenterl}.
We define a random walk on the lattice as a walk
from the origin such that
a walker standing on a site is allowed to move into
one of the eight diagonal directions,
$\pm{\bf e}_x\pm{\bf e}_y\pm{\bf e}_z$.
We require that the walker never pass through
the bonds ${\bf e}_x, {\bf e}_y$
and ${\bf e}_z$.
\begin{figure}
\begin{center}
    \leavevmode
    \epsfysize = 8cm
     \epsfbox{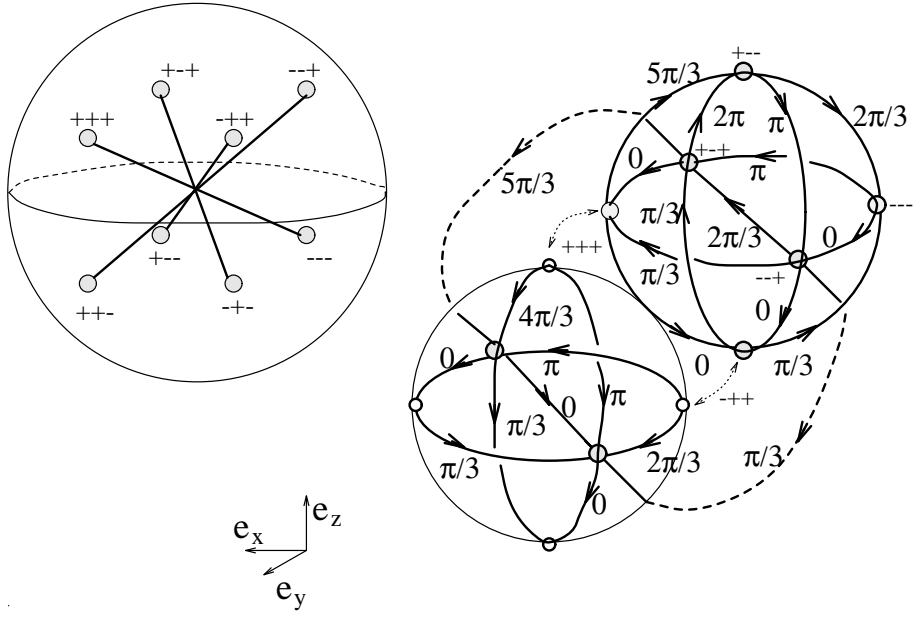}
\end{center}
\caption{Definition of the phase of body centered lattice.}
\label{diagrambody}
\end{figure}
As before, we draw the diagram of a phase assignment
on this lattice as
Fig.\ref{diagrambody},
from which the phase factor, i.e., all the elements of $Q({\bf 0},J)$,
of a walk can be read :
  \begin{equation}
    Q({\bf 0}, J) = \left(
      \matrix{
        1 & e^{\frac 4 3 \pi iJ} & 1 & e^{\frac 5 3 \pi iJ}
        & 0 & e^{-\frac 1 3 \pi iJ} & e^{\pi iJ} & 1 \cr
        e^{-\frac 4 3 \pi iJ} & 1 & e^{-\frac 5 3 \pi iJ} & 1 &
        e^{\frac 1 3 \pi iJ} & 0 & 1 & e^{-\pi iJ} \cr
        1 & e^{\frac 5 3 \pi iJ} & 1 & e^{2\pi iJ} &
        e^{- \pi iJ} & e^{- \frac 2 3 \pi iJ} & 0 & e^{-\frac 1 3 \pi iJ} \cr
        e^{-\frac 5 3 \pi iJ} & 1 & e^{-2 \pi iJ} & 1 &
        e^{\frac 2 3 \pi iJ} & e^{\pi iJ} & e^{\frac 1 3 \pi iJ}& 0 \cr
        0 & e^{-\frac 1 3 \pi iJ} & e^{\pi iJ} & e^{-\frac 2 3 \pi iJ} &
        1 & 1 & 1 & e^{-\frac 1 3 \pi iJ} \cr
        e^{\frac 1 3 \pi iJ} & 0 & e^{\frac 2 3 \pi iJ} & e^{-\pi iJ} &
        1 & 1 & e^{\frac 1 3 \pi iJ} & 1 \cr
        e^{-\pi iJ} & 1 & 0 & e^{-\frac 1 3 \pi iJ} &
        1 & e^{-\frac 1 3 \pi iJ} & 1 & e^{-\frac 2 3 \pi iJ} \cr
        1 & e^{\pi iJ} & e^{\frac 1 3 \pi iJ} & 0 &
        e^{\frac 1 3 \pi iJ} & 1 & e^{\frac 2 3 \pi iJ} & 1 \cr}
        \right)  .
  \end{equation}
The matrix $P({\bf p})$ is determined naturally as
  \begin{equation}
    P({\bf p}) =  \left(
        \matrix{
        e^{i p_{+++}} & & & & & \bigzerou &  &\cr
         & e^{i p_{++-}} & & & & & &\cr
         & & e^{i p_{+-+}} & & & & &\cr
         & & & e^{i p_{+--}} & & & &\cr
         & & & & e^{-i p_{+++}}  & & &\cr
         & & & & & e^{-i p_{++-}} & & \cr
         & & & & & & e^{-i p_{+-+}} & \cr
         & \bigzerol & & & & & & e^{-i p_{+--}} \cr
        }
        \right)
  \end{equation}
where $p_{+++}, p_{++-}, ... $ are defined as
\begin{equation}
       p_{\pm \pm \pm} \equiv \frac 1{\sqrt{3}}(\pm p_x\pm p_y\pm p_z) .
\end{equation}

we see from the form of the matrix $Q({\bf 0},J)$
that the theory has the symmetry
$$J\rightarrow -J, \qquad J\rightarrow J+6 .$$

Using the above data, we define the free energy of the theory.
The eigenvalues of the matrix $Q({\bf 0},J)$ are shown
in the Table \ref{eigenvaluesbc}.
\begin{table}
  \begin{center}
\renewcommand{\arraystretch}{1.2}
   \begin{tabular}{|l||ll|ll|ll|}
    \hline
    {\it Spin Factor} & \multicolumn{6}{c} {\it Eigenvalues}  \vline\\
    \hline \hline
    J = 0  & {\bf 7} & (1) &   1  & (4) &  $-1$  & (3)  \\
    \hline
    J = 1/2 & ${\bf 1+\sqrt{3}+\sqrt{6}}$ & (2) &
              $1+\sqrt{3}-\sqrt{6}$ & (2) & $1-\sqrt{3}$  & (4)  \\
    \hline
    J = 1 & {\bf 4} & (3) & 0 & (3) & $-2$ & (2)  \\
    \hline
    J = 3/2 & ${\bf 1+\sqrt{6}}$ & (4) & $1-\sqrt{6}$ & (4) &  & \\
    \hline
    J = 2 &  ${\bf 4}$ & (2) & 2 & (3) & $-2$ & (3) \\
    \hline
    J = 5/2 & ${\bf 1-\sqrt{3}-\sqrt{6}}$ & (2) &
    $1-\sqrt{3}+\sqrt{6}$ & (2) &
              $1+\sqrt{3}$ & (4) \\
    \hline
    J = 3 & ${\bf -5}$ & (1) &  3 & (3) &  1 & (4) \\
    \hline
  \end{tabular}
  \caption[Table]{{\footnotesize Table of eigenvalues of the
  matrix $Q({\bf 0}, J)$.
Numbers in () represent the degeneracy of the corresponding eigenvalues.}}
 \end{center}
  \label{eigenvaluesbc}
\end{table}

In the case of $J=0$ and  $1/2$, we obtain the same results as those on the
cubic lattice at least
with respect to the free energy of the theory.
Namely, we obtain the same free energy as (\ref{scalingfreeenergyn0}) and
(\ref{scalingfreeenergyn1}) up to the positive coefficient of the mass term.

Since the matrix $Q({\bf p},J)$ is of order eight,
fields with spin higher than $3/2$ might come out.
But the calculation of the eigenvalues of $Q({\bf p},J)$
show that these do not appear.

We can also consider the theory on a
body centered rectangular lattice.
On this lattice,
we also obtain the Dirac spinor field in the $J=\frac 1 2$ case.
Besides, the spin $1$ vector field no more appears
on these deformed lattices in the $J=1$ case as well as on the
triclinic lattice.

%
%
\newsection{Concluding Remarks}
In this paper, we construct spinor field theories
in terms of a particular class of random walks on
several three dimensional lattice structures.
Since we employ such regularization, all physical quantities
can be evaluated as finite ones at any stage
before taking the limit of lattice spacing to be zero.
We showed  the convergence of the infinite summation
over all possible walks.
Therefore, the self-consistency of our construction is clear.

On the regular cubic lattice, two point functions in
the Klein-Gordon, the Dirac and the massive Chern-Simons
theories are obtained for the
corresponding parameters $J=0$, $1/2$ and 1, respectively.
In the $J=1/2$ model there are no fermion doublers.
In the $J=1$ model three other excitations with long range correlation
arise and govern its critical behavior, then the critical
exponents do not satisfy the hyper-scaling relation
but a corresponding inequality.
This is simply due to the two different scale lengths.

What is nontrivial in our results is that no spinor field theory
of spin higher than 1 is obtained.
The situation is unchanged even on the body-centered lattice where
walkers can move to eight-directions from any site.
It should be studied further whether this `$J=1$ barrier' is
a real one or an artifact of our particular choice of
lattice.

We have also argued the universal properties of the construction of
spinor fields on various lattices.
We see that the Klein-Gordon field is obtained on any lattice for $J=0$.

In the case of $J=1/2$, all lattices investigated here permit the
construction of the Dirac spinor field.
A natural expectation is that the Dirac spinor field can be constructed
on any other lattice.

\vspace{5mm}
%
%
The authors thank T.~Hara, K.~R.~Ito, N.~Sakai, G.~Semenoff and H.~Tasaki
for important comments and discussions.
One of them (S.K.) thanks menbers of Yukawa Institute for Theoretical
Physics for kind hospitality extended to him.
They are grateful to J.~Ambj\o rn, P.~Kurzepa and P.~Orland for
notifying them of references.
They thank P.~Crehan for reading the manuscript.
This work is supported in part by Grant-in-Aid for
Scientific Research (S.K.) from the Ministry of Education, Science
and Culture.
\vspace{5mm}
%
%
\def\numberbysectiona{\@addtoreset{equation}{section}
\def\theequation{A.\arabic{equation}}}
\numberbysectiona
\vspace{7mm}
\pagebreak[3]
\setcounter{section}{1}
\setcounter{equation}{0}
\setcounter{subsection}{0}
\setcounter{footnote}{0}
\begin{center}
{\large{\bf Appendix}}
\end{center}
\nopagebreak
\medskip
\nopagebreak
\hspace{3mm}

We prove the convergence property of the free energy defined by
the first equation of the eq.(\ref{freeenergy2}).
We show that the convergence radius of the infinite sum
as a power series in $\kappa$
\be
A(\kappa, J) = \tr \sum^{\infty }_{l=1} \frac 1 l
[\kappa Q({\bf p}, J)]^l
\label{infsum}
\ee
is $|\kappa_c|$, where $\kappa_c{}^{-1}$ is the eigenvalue of
$Q({\bf 0}, J)$ with the largest absolute value.
Let $U({\bf p},J)$ be a unitary matrix which
triangulates the matrix $Q({\bf p},J)$ as
\ba
U({\bf p},J)Q({\bf p},J)U({\bf p},J)^{\dagger}=
\left( \matrix{
\lambda_{1}({\bf p},J) & \cdot & \cdot & \cdot & \cdot & \cdot \cr
 & \lambda_{2}({\bf p},J) & \cdot & \cdot & \cdot & \cdot \cr
 & & \cdot & \cdot & \cdot & \cdot \cr
 & & & \cdot & \cdot & \cdot \cr
 & & & & \cdot & \cdot \cr
\bigzerol & & & & & \lambda_{6}({\bf p},J) \cr
} \right).
\ea
Note that the diagonal elements $\lambda_{i}({\bf p},J) $
$(i=1,...,6)$ are identical to the eigenvalues of the matrix
$Q({\bf p},J)$.
By using this expression, we can deform the sum (\ref{infsum})
as follows :
\ba
A(\kappa, J) &=&
\tr \sum^{\infty}_{l=1} \frac{1}{l}
U({\bf p},J)^{\dagger}
\left[ \kappa U({\bf p},J)Q({\bf p},J)U({\bf p},J)^{\dagger} \right]^l
U({\bf p},J)
\nonumber \\
&=&
\tr \sum^{\infty}_{l=1} \frac{1}{l}
\left[\kappa \left(\matrix{
\lambda_{1}({\bf p},J) &\cdot &\cdot &\cdot &\cdot &\cdot  \cr
& \cdot & \cdot&\cdot &\cdot &\cdot \cr
& & \cdot &\cdot &\cdot &\cdot \cr
& & & \cdot &\cdot &\cdot \cr
& & & & \cdot &\cdot \cr
\bigzerol & & & & & \lambda_{6}({\bf p},J) \cr
}\right)\right]^{l}
\nonumber \\
&=& \lim_{L \rightarrow \infty}
\sum^6_{i=1} \sum^{L}_{l=1} \frac{1}{l}
\left[ \kappa \lambda_i({\bf p},J) \right]^l .
\ea
If $|\kappa \lambda_i({\bf p},J)| < 1$ $(i=1,...,6)$ holds,
this summation converges to be
\be
  \sum^6_{i=1} \log (1 - \kappa \lambda_{i}({\bf p},J)) ,
\ee
which is also represented as
\be
  \log \det \left[ 1 - \kappa Q({\bf p},J) \right]  .
  \label{freeenergydensity3}
\ee

Thus the free energy converges if $\kappa$ is in the region
$|\kappa| < |\kappa_c|$ where $\kappa_c= \kappa_c(J)$ is
the inverse of the eigenvalue whose absolute value is
the largest among the set of eigenvalues of matrices
$Q({\bf p},J)$ for a fixed $J$.

Next, we show that the eigenvalue with the largest absolute value for
a fixed $J$ is given by that for ${\bf p} = {\bf 0}$.
Let $\lambda_i$ $(i=1,...,6)$ be the eigenvalues of $Q(J,{\bf 0})$.
We set that the absolute value of $\lambda_1$ (or $\lambda_6$)
is less (or more) than or equal to
any other $\lambda_i$.
Given a unit vector ${\bf x}$, there exists a unit vector
${\bf y} = (y_1, \cdots, y_6)$
such that
\be
  {\bf x}^{\dagger}\, Q(J,{\bf 0})^{\dagger}Q(J,{\bf 0})\, {\bf x}
  = \sum^6_{i=1}|y_i|^2 \lambda_i{}^2 .
\ee
It follows that
\be
  \lambda_1{}^2 \leq
   {\bf x}^{\dagger}\, Q(J,{\bf 0})^{\dagger}Q(J,{\bf 0})\, {\bf x}
   \leq\lambda_6{}^2 .
  \label{app1}
\ee

For any eigenvalue $\alpha$ of
$Q(J,{\bf p})=Q(J,{\bf 0})P({\bf p})$,
we can choose a unit eigenvector ${\bf x}$ :
$Q(J,{\bf 0})P({\bf p}){\bf x}=\alpha{\bf x}$ .
Then,
\be
  (P{\bf x})^\dagger Q^2 P{\bf x} =|\alpha|^2 .
  \label{app2}
\ee

By applying the formula (\ref{app1}) to (\ref{app2}),
we obtain
\be
  \lambda_1{}^2 \leq |\alpha|^2  \leq \lambda_6{}^2 .
\ee

It has been shown that the eigenvalue of the largest
absolute value is that of ${\bf p} = {\bf 0}$.
Therefore the infinite sum eq.(\ref{infsum}) converges uniformly
with respect to ${\bf p}$ for $|\kappa| < |\kappa_c|$.

%
%

%
\vspace{5mm}
%
\newcommand{\NP}[1]{{\it Nucl.\ Phys.\ }{\bf #1}}
\newcommand{\PL}[1]{{\it Phys.\ Lett.\ }{\bf #1}}
\newcommand{\CMP}[1]{{\it Commun.\ Math.\ Phys.\ }{\bf #1}}
\newcommand{\MPL}[1]{{\it Mod.\ Phys.\ Lett.\ }{\bf #1}}
\newcommand{\IJMP}[1]{{\it Int.\ J. Mod.\ Phys.\ }{\bf #1}}
\newcommand{\PR}[1]{{\it Phys.\ Rev.\ }{\bf #1}}
\newcommand{\PRL}[1]{{\it Phys.\ Rev.\ Lett.\ }{\bf #1}}
\newcommand{\PTP}[1]{{\it Prog.\ Theor.\ Phys.\ }{\bf #1}}
\newcommand{\PTPS}[1]{{\it Prog.\ Theor.\ Phys.\ Suppl.\ }{\bf #1}}
\newcommand{\AP}[1]{{\it Ann.\ Phys.\ }{\bf #1}}
\newcommand{\JP}[1]{{\it J.\ Phys.\ }{\bf #1}}
\newcommand{\JETP}[1]{{\it JETP \ }{\bf #1}}
\newcommand{\AJP}[1]{{\it Am.\ J.\ Phys.\ }{\bf #1}}
\newcommand{\JMP}[1]{{\it J.\ Math.\ Phys.\ }{\bf #1}}
%


%
\end{document}